\documentclass[twocolumn,showpacs,amsmath,amssymb,prd]{revtex4}

\usepackage{graphicx}
\usepackage{dcolumn}
\usepackage{bm}
                
\def\be{\begin{equation}}
\def\ee{\end{equation}}  
\def\ba{\begin{eqnarray}}
\def\ea{\end{eqnarray}}
\def\<{\langle}
\def\>{\rangle}

\begin{document}
\input{epsf}

\title{Atmospheric neutrinos in ice and measurement of neutrino oscillation parameters}
\author{Enrique Fernandez-Martinez$^1$, Gerardo Giordano$^2$, Olga Mena$^{3}$, Irina Mocioiu$^2$}
\affiliation{$^1$ Max-Planck-Institut f\"{u}r Physik (Werner-Heisenberg-Institut), 
F\"{o}hringer Ring 6, 80805 M\"{u}nchen, Germany}
\affiliation{$^2$ Department of Physics, 
Pennsylvania State University, University Park, PA 16802, USA}
\affiliation{$^3$ 
Instituto de F\'{\i}sica Corpuscular, IFIC, CSIC and Universidad de Valencia, Spain}

\begin{abstract}
The main goal of the IceCube Deep Core Array is to search for 
neutrinos of astrophysical origins. Atmospheric neutrinos are commonly considered as a background for these searches. We show 
that the very high statistics atmospheric neutrino data can be used to obtain precise measurements of the main oscillation parameters.
 \end{abstract}


\date{\today}
\maketitle

\section{Introduction}

A large number of experiments of different types have provided strong evidence for neutrino oscillations and thus for physics beyond the Standard Model (see Ref.~\cite{GonzalezGarcia:2007ib} and references therein). 
 
Cosmic ray interactions in the atmosphere give a natural source of neutrinos. These atmospheric neutrinos in the GeV range have been
used by the Super-Kamiokande (SK) detector to provide evidence for neutrino oscillations~\cite{Ashie:2005ik}. The large size of neutrino telescopes such as AMANDA, IceCube and KM3NeT makes possible the detection of a large number of atmospheric neutrino events with a higher energy threshold, $\sim 100$~GeV, even though the neutrino flux decreases rapidly with energy ($\sim E_\nu^{-3}$). These observatories were built to detect neutrinos from astrophysical sources, or from the annihilation of Weakly Interacting Massive Particles (WIMPs)~\cite{telescopes}. Because of the high threshold energies, neutrino oscillation effects in these ice/water Cherenkov detectors are small. 

Recently, a low energy extension of the IceCube detector, 
the IceCube Deep Core array (ICDC) has been proposed and deployed~\cite{deepcore}. It consists of six densely instrumented strings ($7{\rm m}$ spacing between optical modules) located in the deep center region of the IceCube detector plus the seven nearest standard IceCube strings. Its goal is to significantly improve the
atmospheric muon rejection and to extend the IceCube neutrino
detection capabilities in the low energy domain, down to muon or cascade energies as low as $5$~GeV. The instrumented volume is $15$~Mton. Such a low threshold array bubried deep inside IceCube will open up a new energy window on the universe. It will search for neutrinos from sources in the Southern hemisphere, in particular, from the galactic center region, as well as for neutrinos from WIMP annihilation, as originally motivated. 
In \cite{MMR} we have proposed neutrino oscillation physics as a further motivation for building such an array. In particular, we have analyzed the sensitivity of ICDC to the neutrino mass hierarchy. 
ICDC can detect tens of thousands of atmospheric neutrino events per year, 
orders of magnitude beyond the present data sample, providing rich 
opportunities for detailed oscillation studies. In \cite{GMM} we have also analyzed the cascade signal in ICDC, showing that this can be used to get a high rate of $\nu_\tau$ interactions. 

Here we concentrate on the muon neutrino disappearance analysis in ICDC, exploring the precision that can be reached in the measurement of the main neutrino oscillation parameters, $\Delta m^2_{31}$ and $\theta_{23}$.

Section \ref{sec:osc} reviews briefly our present knowledge of the neutrino oscillation parameters as well as their expected uncertainties from near future facilities. We then proceed in section \ref{sec:results} to describe the analysis presented here and the results obtained from it. We close with an outlook in section \ref{sec:outlook}. 

\section{Neutrino Oscillations}
\label{sec:osc}

Neutrino data from solar, atmospheric, reactor and accelerator
experiments is well understood in terms of three-flavor neutrino
oscillations. Two $\Delta m^2$ values and two (large) mixing angles
are well determined, while the third mixing angle is limited to be
very small. The CP-violating phase ($\delta$) is completely
unconstrained. In addition, the sign of $\Delta m^2_{31}$ is also
unknown.

The best fit {oscillation} parameter values obtained from present data
are \cite{GonzalezGarcia:2010er}:

\ba
|\Delta m^2_{31}|&=&2.56\times 10^{-3} {\rm eV}^2\nonumber\\
\Delta m^2_{21}&=& 7.6 \times 10^{-5} {\rm eV}^2\nonumber\\
\sin^22\theta_{23}&=& 0.99\nonumber\\
\tan^2\theta_{12}&=& 0.47
\ea
and $\sin^22\theta_{13}\le 0.15$ for $\Delta m^2_{31}=2.5\times 10^{-3}
{\rm eV}^2$. Notice that an extra unknown in the neutrino oscillation
scenario is the octant in which $\theta_{23}$ lies, if $\sin^2 2
\theta_{23}\ne 1$. This has been dubbed in the literature as ``the
$\theta_{23}$ octant ambiguity". At the three sigma level $|\Delta m^2_{31}|$ can vary between $2.0 - 2.83 \times 10^{-3} {\rm eV}^2$, while $\theta_{23}$ varies between $35.5^{\circ}-53.5^{\circ}$.

In the near future, long baseline experiments like MINOS~\cite{minos} and T2K \cite{T2K} will improve the current precision on $\Delta m^2_{31}$ and possibly discover a non-zero value of $\theta_{13}$, if this is close to the present upper limit. In a few years, reactor experiments like DoubleChooz~\cite{dchooz}, RENO~\cite{reno} and Daya Bay~\cite{dayabay} will provide improved sensitivity to $\theta_{13}$. This information can be used as input in our analysis,
reducing some of the parameter uncertainties.

In the past, atmospheric neutrinos in the Super-Kamiokande detector
have shown evidence for neutrino oscillations and the first
measurements of $|\Delta m^2_{31}|$ and $\sin^22\theta_{23}$~\cite{Ashie:2005ik}, providing compelling evidence for neutrino oscillations versus more exotic phenomena~\cite{exotic}. While facing more systematics than accelerator/reactor experiments due to the uncertainties in the (natural source) neutrino fluxes, atmospheric neutrinos provide great opportunities for exploring oscillation physics due to the large range of energies and pathlengths that they span. The ICDC detector 
will collect a data sample which is a few orders of magnitude larger than that of the Super-Kamiokande experiment and can also measure energy and directional information, such that many of the systematic errors associated 
with unknown normalizations of fluxes, cross-sections, etc. can be much better understood and reduced by using the data itself. The MINERVA \cite{MINERVA} experiment will also provide important information about cross-sections in the relevant energy range.

The downward going neutrinos are largely unaffected by oscillations, so they can be used for determining the atmospheric neutrino flux.

In our numerical calculations we have taken into account full three flavor oscillations. It is however straightforward to see that solar parameters do not play an important role in the analysis due to the rather high energy threshold of ICDC.

\section{Analysis and Results}
\label{sec:results}

We concentrate our analysis on the very high statistics track signal. This is dominated by muon events originating from the charged current (CC) interactions of muon neutrinos. Additional small contributions come from $\nu_e\to\nu_\mu$ oscillations for non-zero values of $\theta_{13}$, as well as $\nu_\mu\to\nu_\tau\to\tau\to\mu$ events. At low energies cascade events from various interactions of all neutrino flavors might not be distinguished from tracks, so they also contribute to the same event rate.

We investigate the neutrino energy range between $10$~GeV and $100$~GeV, assuming bins of $5$~GeV width in the {\it observable} muon energy. Note that the energy of secondary muons from CC interaction in the $10-100$ GeV neutrino energy range of interest here is $\langle E_\mu \rangle = 0.52 E_\nu$ and $\langle E_{\bar\mu}\rangle = 0.66 E_{\bar\nu}$, respectively for neutrinos and antineutrinos \cite{GQRS}. For our numerical calculations, we use differential neutrino interaction cross-sections as given by \cite{GQRS} with CTEQ6 parton distribution functions. 

As previously mentioned, we only include upward-going neutrinos in the analysis for determining the neutrino oscillation parameters. The down-going neutrino events are not sensitive to oscillations, but could improve uncertainties in the normalization of the atmospheric neutrino flux. 

We evaluate the number of muon events from muon neutrino interactions as a function of observable energy and zenith angle as:
\begin{eqnarray}
& & N_{i,j,\mu} (E_\mu, c_\nu)={2 \pi N_A \rho \, t} \, 
\int_{E_i}^{E_i +\Delta_i}dE_\mu \int_{c_{\nu, j}}^{c_{\nu,j}
  +\Delta_j} dc_\nu 
  \nonumber\\
& & \int_{E_\mu}^{\infty}dE_\nu V_\mu
\frac{d\phi_{\nu_\mu}}{dE_\nu d\Omega}(E_\nu,c_\nu)\, \frac{d\sigma^{\rm  CC}_{\nu_\mu}}{d E_\nu} (E_\nu,E_\mu) P_{\nu_\mu \to \nu_\mu}(E_\nu, c_\nu)\nonumber\\
&  &+ \nu\to\bar\nu
~.
\label{eq:events}
\end{eqnarray}
The other contributions to the event rate are obtained in a similar way, using the corresponding initial fluxes, oscillation probabilities and cross-sections.

In the above equation, $\Delta_i$ and $\Delta_j$ are, respectively, the bin widths of the $i$ energy bin and $j$ zenith angular bin (defined by $c_\nu=\cos\theta_{zenith}$), $t$ is the exposure time, $d\phi_\nu$'s are the atmospheric (anti)neutrino differential spectra and $d \sigma^{\rm CC}/d E$ are differential CC (anti)neutrino cross sections. 

The oscillation probabilities $P_{\nu_\alpha\to\nu_\beta}$ have been obtained in a full three flavor scenario, including matter effects. While the signal is dominated by muon disappearance, given the very high statistics, sub-dominant effects become observable if $\theta_{13}$ is within reach of the near future reactor and accelerator experiments.

For the atmospheric (anti)neutrino fluxes, ${d\phi_{\nu_\alpha}}/{dE_\nu d\Omega}$ we use the results from Refs.~\cite{fluxesus}. The atmospheric neutrino fluxes from Refs.~\cite{hondaetal} give similar results.
The absolute electron and muon atmospheric (anti)neutrino
fluxes have errors of $10\%-15\%$ in the energy region of
interest here~\cite{uncertainties}. Those errors are mostly induced by
our incomplete understanding of hadron production, although the situation is expected to improve with HARP and MIPP data. In addition, a
calibration of this overall systematic uncertainty can be performed by looking at different angular apertures/energy ranges where the oscillatory signal is not present, For the neutrino-antineutrino flavor ratios $\phi_{\nu_\mu}/\phi_{\bar\nu_\mu}$ and $\phi_{\nu_e}/\phi_{\bar\nu_e}$ the 
uncertainty is reduced to $\sim 7\%$ in the energy range we explore in the current study~\footnote{In general, the different available computations of the 
atmospheric neutrino fluxes~\cite{fluxesus,hondaetal} predict almost 
the same neutrino-antineutrino ratios.}. 
Even smaller uncertainties are expected when the
muon-to-electron flavor ratio $(\phi_{\nu_\mu} + \phi_{\bar\nu_\mu})/(\phi_{\nu_e}+\phi_{\bar\nu_e})$ is considered. We will comment on the impact of the atmospheric neutrino flux uncertainties on our results below, when we discuss systematic uncertainties in our numerical analysis.

The factor $V_{\mu}$ in Eq. \ref{eq:events} accounts for the effective detector volume. Unlike in many other detectors that use fixed geometric cuts and sharp detection thresholds, neutrino telescopes employ an energy-dependent effective volume, which depends on the type of observable considered, in order to maximize the number of events in their data analysis. Preliminary ICDC estimates \cite{deepcore} show that this effective volume is of the order of 10~Mtons for 10~GeV neutrinos, increasing with energy.  We present here results for our own effective volume estimates, based on several different physics-motivated assumptions regarding the detector. An optimistic configuration consists in using the full $\sim$ 15~Mtons detector mass, corresponding to the 250~m diameter, 350~m height detector. The conservative option that we have adopted is to consider only a 35~m radius around each of the 6 densely instrumented strings (at the ICDC depth, the light attenuation length is 40-45m). This reduces the total mass to $\sim$ 7~Mtons, but avoids overestimating the observable low energy events, which are crucial for the precise determination of oscillation parameters, as we will further discuss below. 

For a fixed detector configuration, we considered two different scenarios to define the effective volume: (1) allow only fully contained events, and (2) allow some of the muons exit the detector. In both cases we require the initial neutrino interaction occurs inside the detector volume. Note that ICDC uses the large IceCube detector as veto, in order to exclude events that originate outside the ICDC region. For the fully contained events, for a detector with cylindrical shape of radius $r$ and height
$h$, $V_\mu$ is given
by~\cite{AS01}
\begin{widetext}
\begin{equation}
V_\mu(E_\mu,\theta)= 2 h r^2
\arcsin\left(\sqrt{1-\frac{R^2_\mu(E_\mu)}{4 \, r^2}\sin^2 \theta}
\right) \left(1-\frac{R_\mu(E_\mu)}{h}|\cos\theta|\right)~,
\label{eq:range}
\end{equation}
\end{widetext}
where $R_\mu(E_\mu)$ is the energy-dependent muon range in ice. 
The fully contained events would yield a better energy measurement. On the other hand, the atmospheric neutrino flux decreases very fast with energy and retaining muons that exit the detector helps improve the statistics at higher energies. These two possibilities turn out to give equivalent results in the final analysis, as most of the information about oscillation parameters is contained in the low energy data, where all events are fully contained without imposing further constraints. 

We always require the neutrino interaction to occur inside the detector volume. ICDC aims to detect the cascade from the initial neutrino interaction in addition to the muon track, which would provide a better measurement of the total initial neutrino energy. Availability of such information would improve our analysis, which only relies on the measurement of the muon energy. 

Since no detailed ICDC simulations of the expected systematic uncertainties is presently available, we discuss results for different assumptions regarding these
uncertainties. 
Systematics that affect the detector observables include angular reconstruction, muon track length reconstruction (directly related to the energy measurement), uncertainties introduced by the modeling of light propagation and light detection efficiency of the optical modules. The uncertainty on the energy and angular measurements is related to the estimation of the effective area/volume, which is a function of both the muon energy and the muon direction. As previously mentioned, ICDC does not have a sharp energy threshold and the ICDC analysis shows that the detector can trigger muons with energies as low 1~GeV and reach an effective volume of 10~Mtons for 10GeV neutrinos. We use a threshold of 5G~eV muon energy in our analysis. Given the detector properties, we estimate the energy resolution to be given by a 25m track length uncertainty, which translates into about 5~GeV muon energy. 

We explore the sensitivity of the atmospheric neutrino data to the main oscillation parameters, $\Delta m^2_{31}$ and $\theta_{23}$ using a $\chi^2$ statistical analysis, which contains both statistical and systematic uncertainties. 
To account for systematic uncertainties we have included an overall error computed as a fraction $f_{sys}$ of the number of events in each bin: $f_{sys}\cdot N_{i,j,\mu} (E_\mu, c_\nu)$, added in quadrature with the statistical error. We will consider $f_{sys}$ values of 5\% and 10\% for our analysis.
This treatment of systematic uncertainties is more conservative than other more elaborated approaches used in analysis of neutrino data, as, for instance, the so-called pull method~\cite{Huber:2002mx,Fogli:2002pt}. This systematic uncertainty can account for the issues discussed above, including uncertainties in fluxes, cross-sections and total detector volume. Given that many of the uncertainties discussed can be reduced once data becomes available and the angular and energy regions not sensitive to the oscillation signal are used, a 10\% overall systematic uncertainty is likely an overestimate of what can be achieved. Other types of uncertainties, concerning angular resolution, energy resolution and energy thresholds are explored in more detail below, as we consider different scenarios in our analysis.

\begin{figure*}[t]
\includegraphics[width=2.32in]{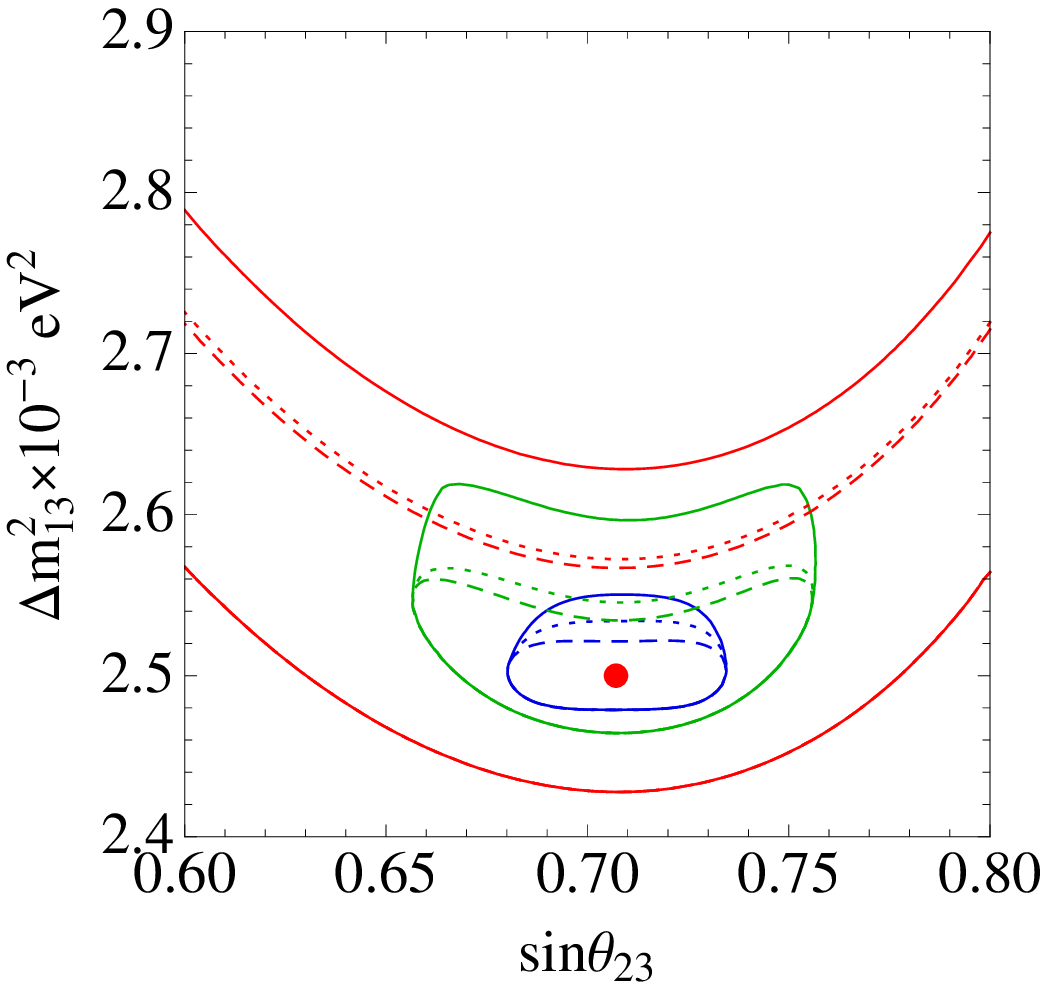}
\includegraphics[width=2.32in]{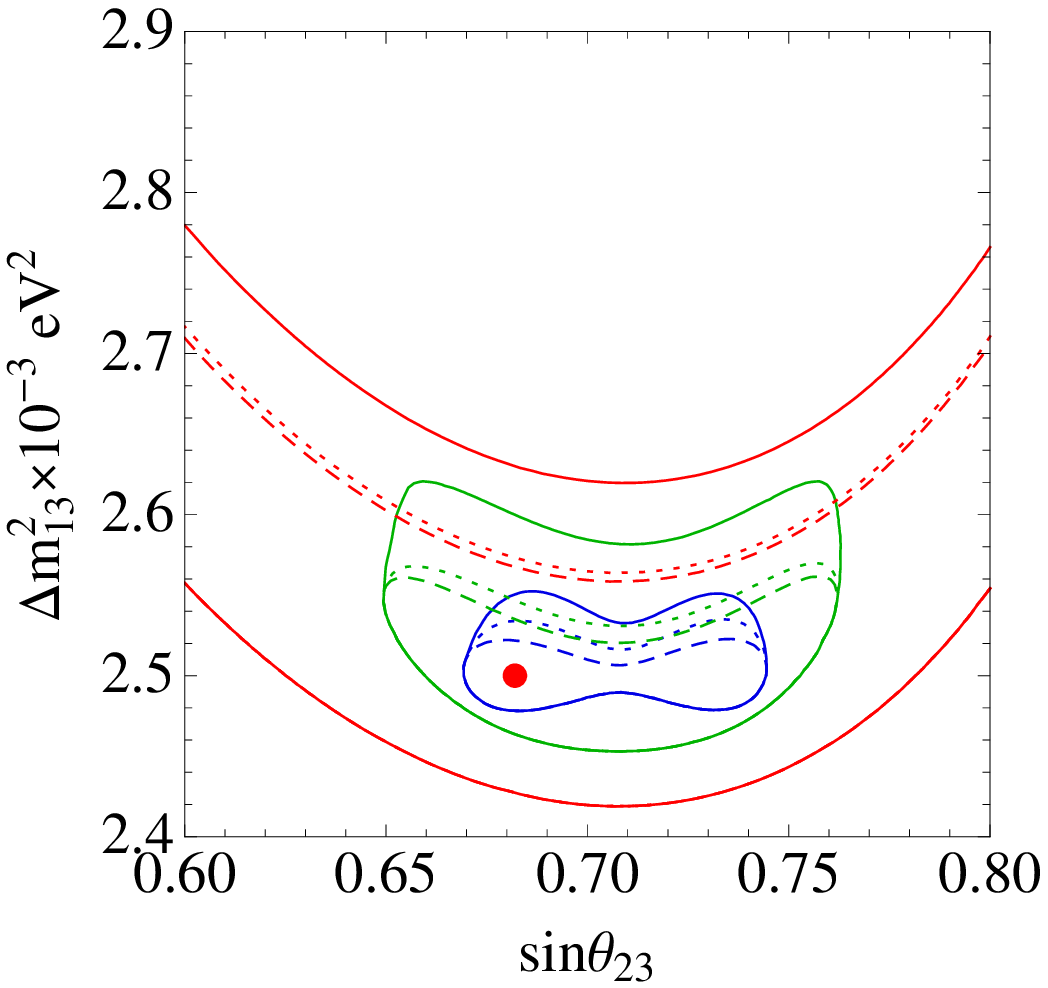}
\includegraphics[width=2.32in]{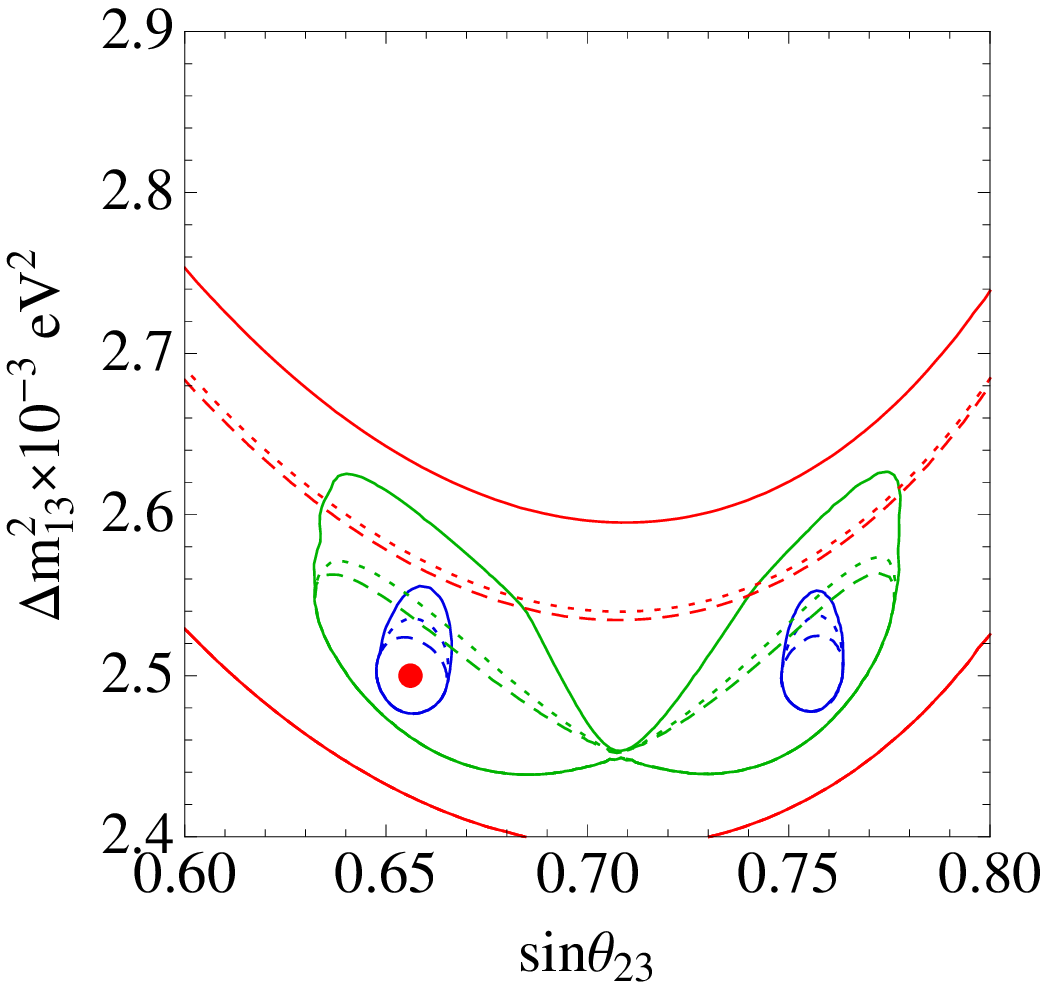}
\caption{The left, middle and right panels show the 90\% CL contours for 2 dof, for atmospheric mixing angles $\theta_{23}= 45^\circ, 43^\circ \,{\rm and}$ $41^\circ$, respectively. The different colors account for different energy/angular resolutions.  The solid lines allow $\theta_{13}$ varying freely in its presently allowed range. The dashed lines show the results for a fixed $\theta_{13}$ at the assumed true value $\sin^22\theta_{13}=0.01$. The dotted lines corresponds to free $\theta_{13}$, but with a gaussian prior centered around the true value $\sin^22\theta_{13}=0.01$, with a 1 sigma uncertainty of 0.02. A systematic uncertainty of 5$\%$ has been included.}
\label{fig:contours01}
\end{figure*}

Figure \ref{fig:contours01} shows the 90\% CL contours for the allowed parameter values for two degrees of freedom (dof), with the atmospheric neutrino data collected in ICDC after a 10 years exposure.  Note that our choice of 2 dof statistics is rather conservative as explained in the Appendix of Ref.~\cite{stat}.
The three plots correspond to three different reference values of $\theta_{23}: 45^\circ, 43^\circ \,{\rm and}\, 41^\circ$, all within the 1 sigma presently allowed region.  The very large statistics, of up to 10,000 events per year, provide very good sensitivity to oscillation physics and it can be seen that the precision will be significantly improved compared with the present one. Maximal mixing could be potentially excluded for values of $\theta_{23}$ below $42^\circ$. 

The final sensitivity to the parameters of interest is however strongly influenced by our knowledge of $\theta_{13}$ and various types of systematic uncertainties. We explore these effects in some detail and show that under reasonable assumptions oscillation parameters can be measured with much higher precision than present data allows. 

For Figure \ref{fig:contours01} we have used a reference value of $\sin^2 2 \theta_{13}=0.01$. In order to understand the influence of this angle on the final sensitivity, we show results for three different assumptions, represented by different types of lines. With our present knowledge of $\theta_{13}$ the results would be those represented by solid lines, which correspond to $\theta_{13}$ varying freely in its presently allowed range. The dashed lines show the results for the unrealistic case of a fixed $\theta_{13}$ at the assumed true value. The dotted lines corresponds to free $\theta_{13}$, but with a gaussian prior centered around the true value and with a 1 sigma uncertainty of 0.02. This is a very realistic situation, which assumes having an independent measurement of $\theta_{13}$ from a different experiment, which is expected on the relevant time scale. DoubleChooz should reach an uncertainty around 0.03 and Daya Bay around 0.01 or even smaller, which would improve this analysis even further. It can be seen, however, that even with our conservative assumption, the sensitivity is very close to the ideal situation of fixed $\theta_{13}$.

The different colors on the same plot depict the effects of the energy/angular resolution. The blue lines assume 5 GeV muon energy bins and four different angular bins: $c_\nu\in (-1, -0.9), (-0.9, -0.8), (-0.8, -0.7)$ and $(-0.7, 0)$. The green lines assume that in the lowest energy bins directional reconstruction is very hard and a single angular bin of  $c_\nu\in (-1, 0)$ is considered. The red lines correspond to eliminating the two lowest energy bins from the data, thus setting the neutrino energy threshold around 30 GeV. 

It can be seen that poor angular resolution deteriorates the results somewhat, but the energy threshold is crucial for a good measurement of $\theta_{23}$. For astrophysical source searches, a 30~GeV energy threshold would be sufficient and potentially even desired, in order to reduce the atmospheric neutrino background. The detector setup allows however for a much lower energy threshold, with an estimated effective volume of the order of 10~MTon at neutrino energies of 10~GeV. The oscillation effects become very small above 40~GeV so it is important to analyze the data at the lowest energies available in order to obtain the maximum amaount of information about neutrino oscillation parameters. It is also very important for the measurement of atmospheric neutrino parameters to have events with energies both above and below the oscillation peak and have information about the shape of the energy spectrum, which is most sensitive to the oscillation parameters. For bins with energy above the peak, the allowed regions in the $(\sin\theta_{23},\Delta m^2_{31})$ plane curve upwards, while for energies smaller than the peak they curve downwards. The combination of the two regions has a small intersection and provides a very good measurement. For higher energy threshold, for most pathlength, only the events above the peak would be observed and the sensitivity to $\theta_{23}$ is greatly degraded. 

In addition to the effects of $\theta_{13}$, energy and angular resolution effects,  we have also added a 5\% systematic error in the above results to account for flux/cross-section and other uncertainties, as previously discussed. This uncertainty can be greatly reduced by using the data itself to fit for a flux parameter and by including down-going and higher energy events.

\begin{figure*}[t]
\includegraphics[width=2.32in]{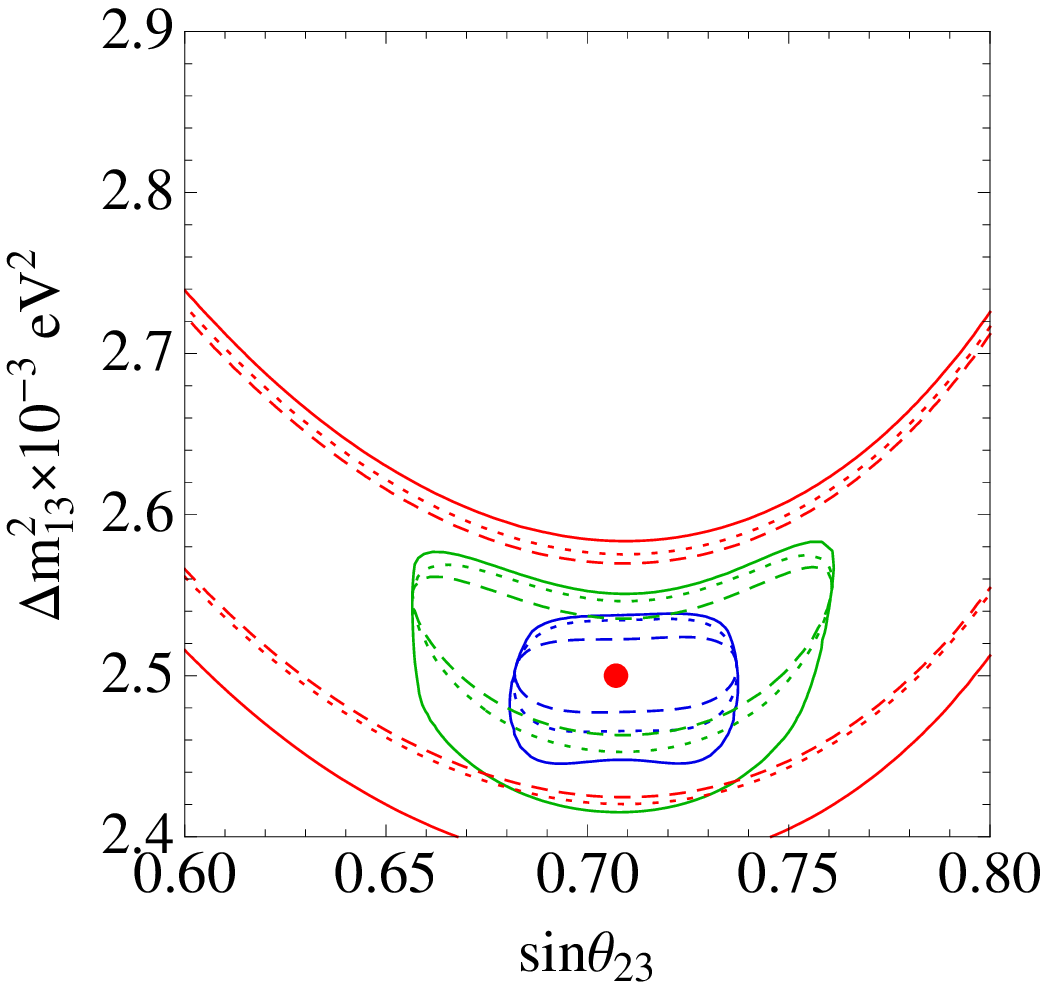}
\includegraphics[width=2.32in]{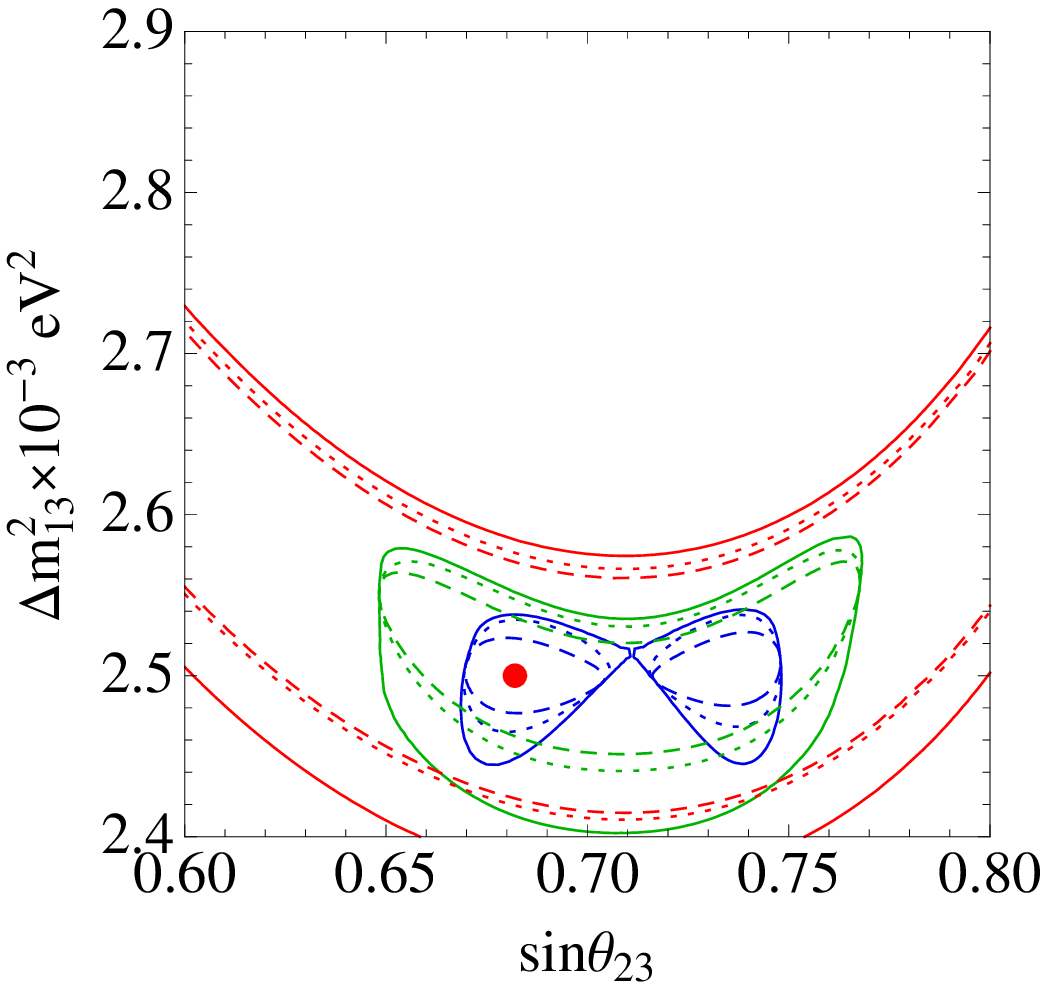}
\includegraphics[width=2.32in]{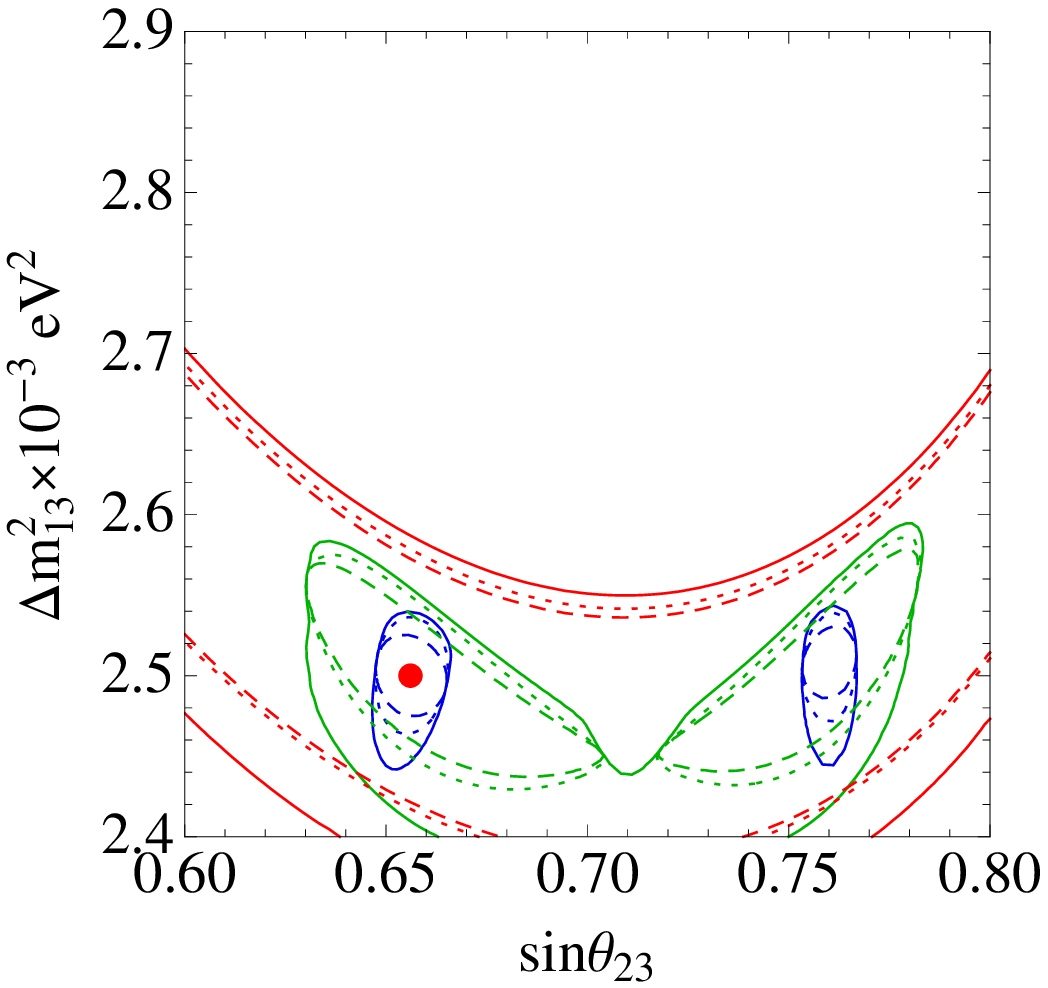}
\caption{Same as Fig.\ref{fig:contours01} but assuming $\theta_{13}$ to be centered at $\sin^2 2 \theta_{13}=0.08$.}
\label{fig:contours08}
\end{figure*}

Figure \ref{fig:contours08} shows the equivalent to Fig.~\ref{fig:contours01} but with a reference value of $\sin^2 2 \theta_{13}=0.08$.
For the larger value of $\theta_{13}$ illustrated in Fig.~\ref{fig:contours08}, the contours become slightly asymmetric, as expected. This asymmetry in $\theta_{23}$ could in principle be used to get sensitivity to the octant. Our analysis shows that it may be possible to determine the octant for small $\theta_{23}$, large $\theta_{13}$ and favorable assumptions about energy/angular resolution and systematics errors. For $\sin^2 2\theta_{13}=0.08$ there is sensitivity to the octant at the 90\% CL if $\theta_{23}$ is smaller than $41^\circ$, while for $\sin^2 2\theta_{13}=0.1$ it might be possible to have octant sensitivity for $\theta_{23}$ smaller than $43^\circ$. These values are still in the 1 sigma presently allowed regions for the atmospheric mixing angle $\theta_{23}$. The octant degeneracy~\cite{Fogli:1996pv} is very hard to resolve and future neutrino facilities such as neutrino factories and/or superbeams~\cite{BurguetCastell:2002qx,Donini:2005db,Mena:2005ek,Geer:2007kn,Meloni:2008it}, the combination of superbeams with atmospheric neutrino data~\cite{Huber:2005ep,Campagne:2006yx}, or future iron calorimeter atmospheric neutrino detectors~\cite{Choubey:2005zy} have been shown to provide powerful setups to resolve it. High statistics atmospheric neutrino data in ICDC could be a first step in addressing this degeneracy, if $\theta_{13}$ turns out to be large and $\theta_{23}$ non-maximal.

It is important to note that a three flavor analysis considering the phyical quantity $\sin\theta_{23}$ can yield much better sensitivity than a two flavor one in terms of $\sin^2 2\theta_{23}$. In Ref. \cite{Minakata:2004pg} it has been shown that going from $\sin^2 2\theta_{23}$ to $\sin\theta_{23}$ one can introduce errors of $10-20\%$ for $\theta_{23}$ values close to $\pi/4$. 

All the above plots assume a reference value of $\Delta m^2_{31}=2.5 \times 10^{-3} {\rm eV}^2$. Changing the reference value of $\Delta m^2_{31}$ essentially moves the allowed regions up and down, without a significant change in the size and shape of the contours. 

Since the track signal is dominated by the muon neutrino disappearance channel, a non-zero CP-violating phase has a very small effect on the parameter sensitivity.
For the largest values of $\theta_{13}$, the effect of the phase would be to slightly enlarge the $\Delta m^2_{31}$ allowed range towards lower values. 

Considering smaller values for $\theta_{23}$ within the 1, 2 or 3 sigma presently allowed regions makes a better measurement even easier, allowing for both the exclusion of maximal mixing and, for relatively large values of $\theta_{13}$, for octant discrimination. We have not shown values of $\theta_{23}$ larger than $45^\circ$, as the results for $\theta_{23}=\pi/4+\delta$ look like ``mirror images" of those for $\theta_{23}=\pi/4-\delta$.

\begin{figure*}[t]
\includegraphics[width=2.32in]{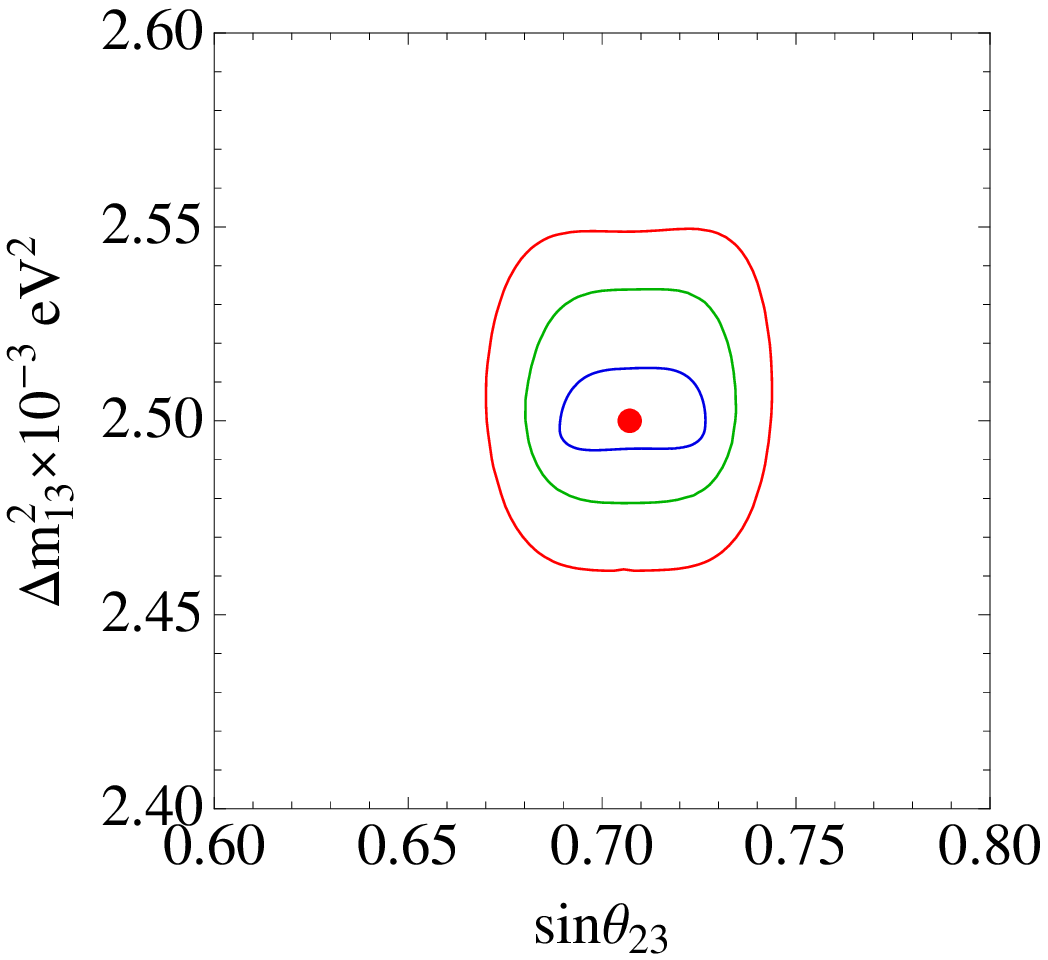}
\includegraphics[width=2.32in]{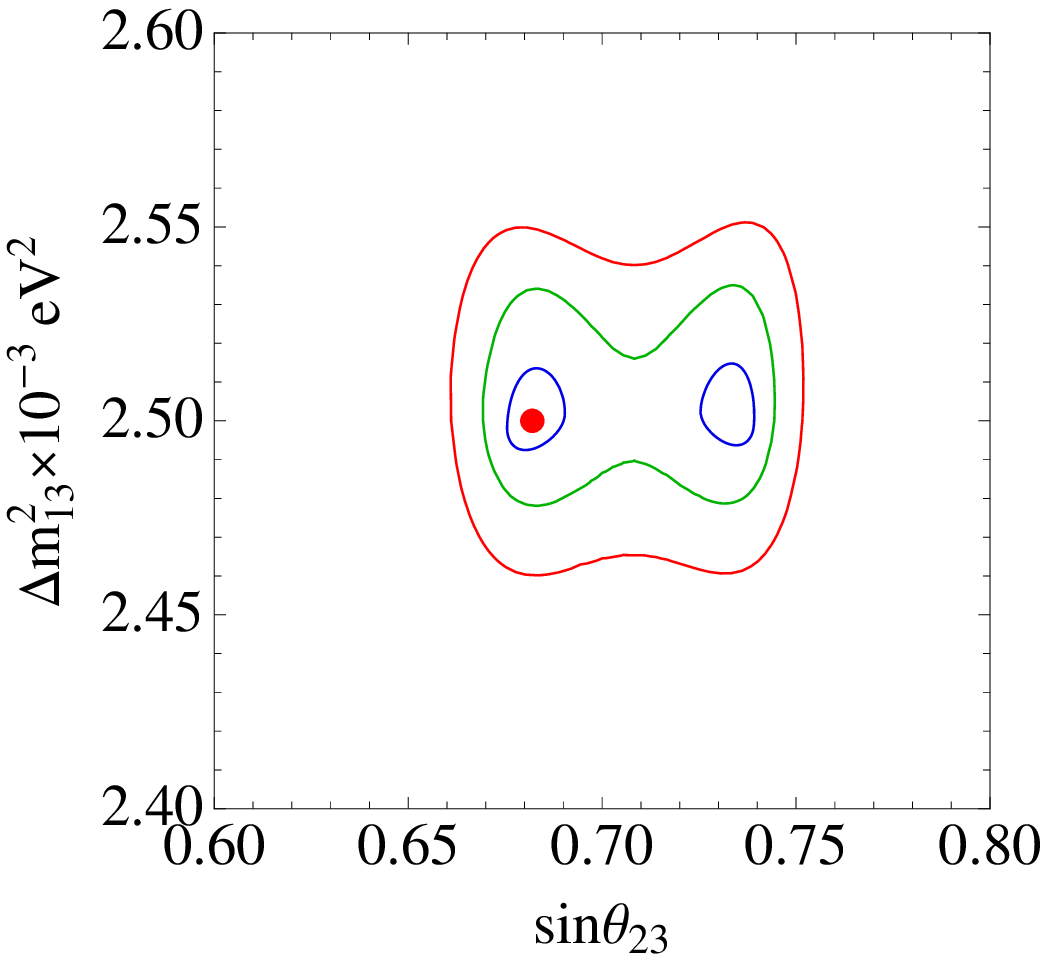}
\includegraphics[width=2.32in]{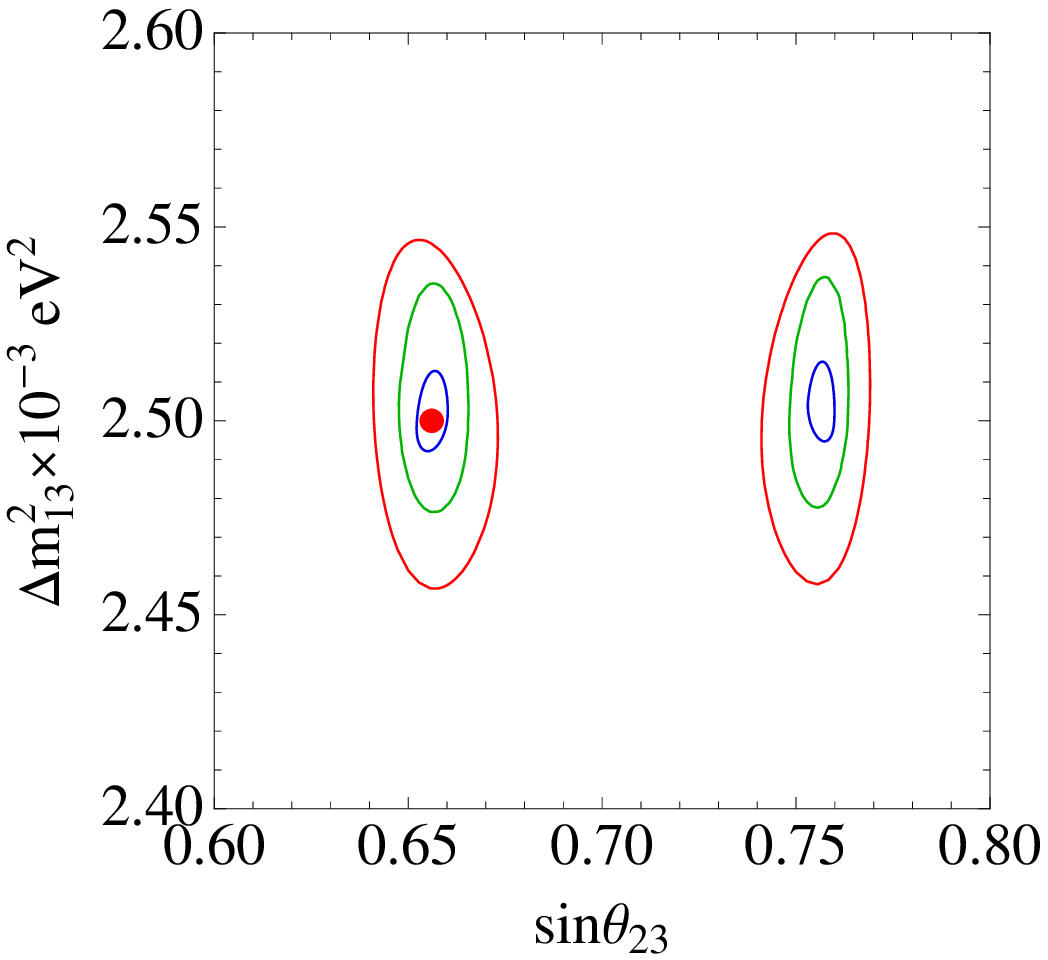}
\caption{The left, middle and right panels show the 90\% CL contours for 2 dof assuming  $\theta_{13}=0.01$ and $\theta_{23}= 45^\circ, 43^\circ \,{\rm and}$ and $41^\circ$, respectively. The different colors account for different values of the systematic uncertainties, see the text for details. We have assumed free $\theta_{13}$ with a gaussian prior centered around the true value $\sin^22\theta_{13}=0.01$, with a 1 sigma uncertainty of 0.02.}
\label{fig:contours_sys01}
\end{figure*}

\begin{figure*}[t]
\includegraphics[width=2.32in]{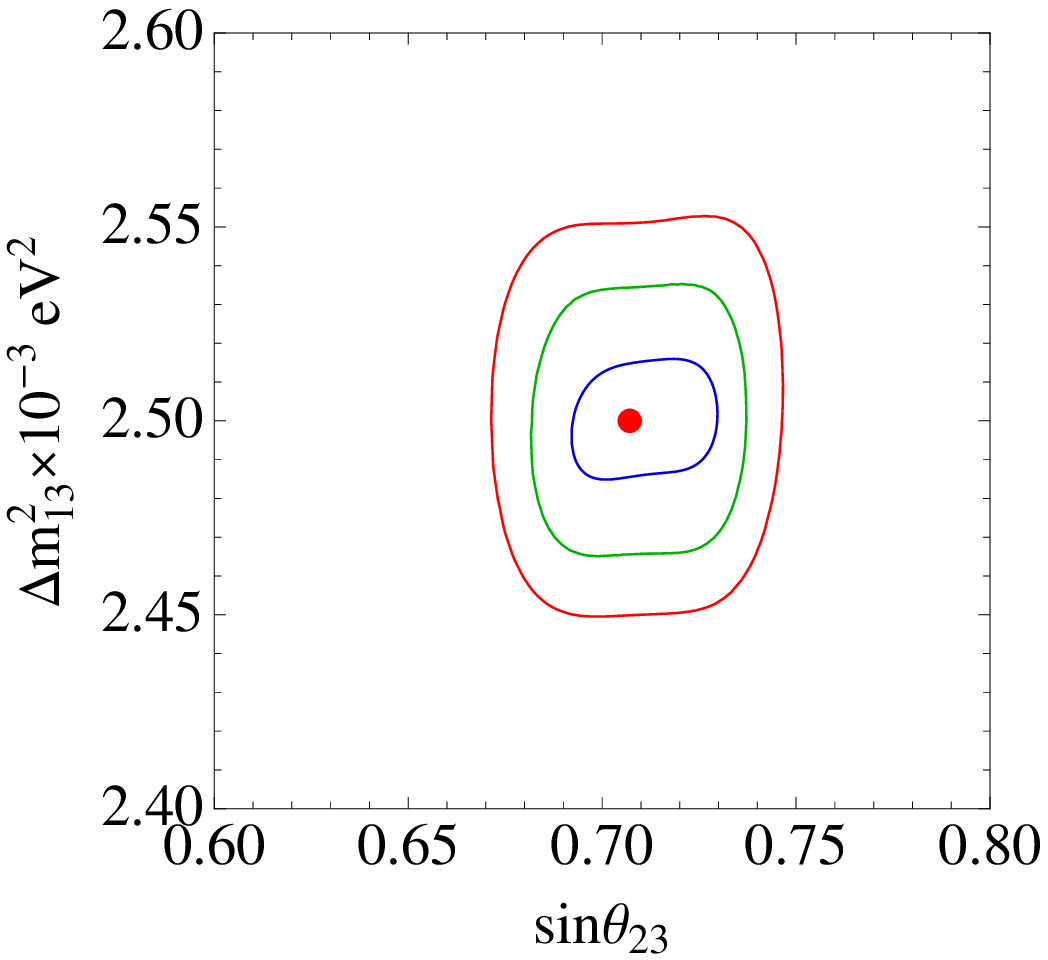}
\includegraphics[width=2.32in]{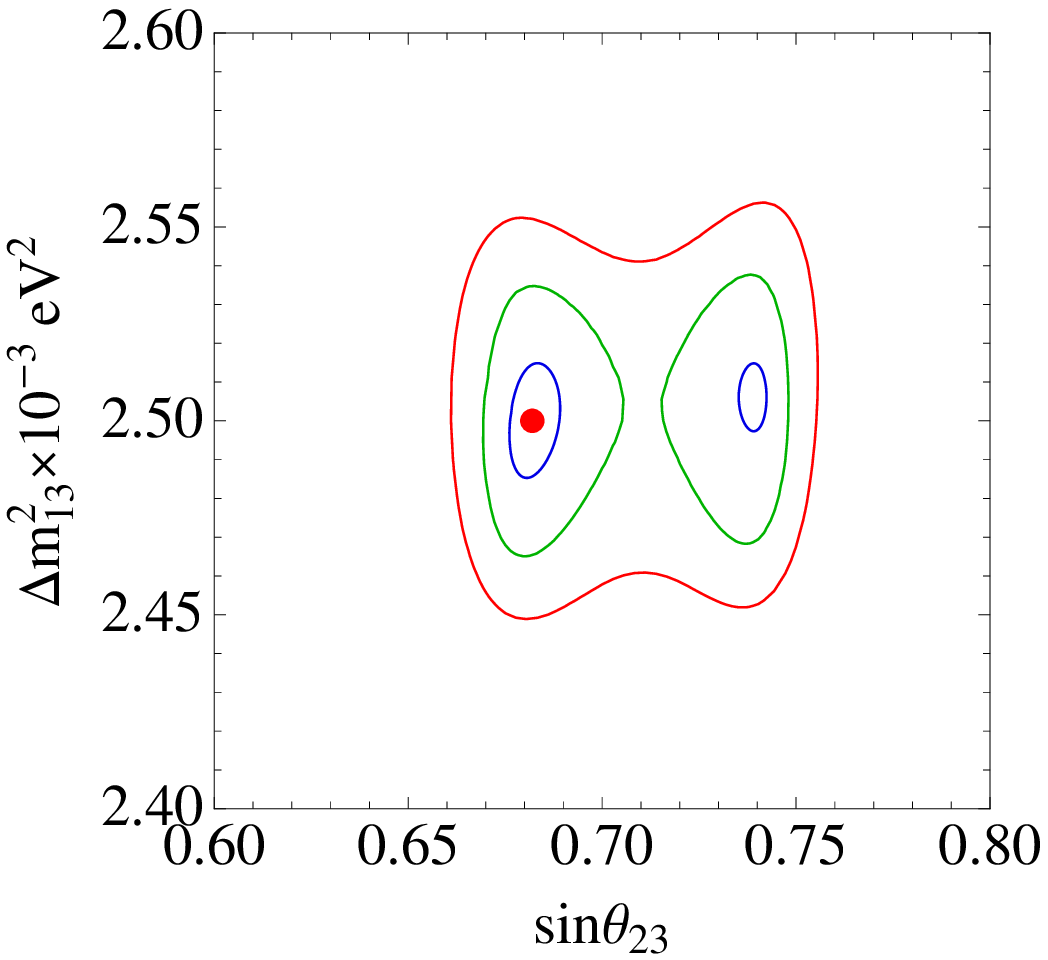}
\includegraphics[width=2.32in]{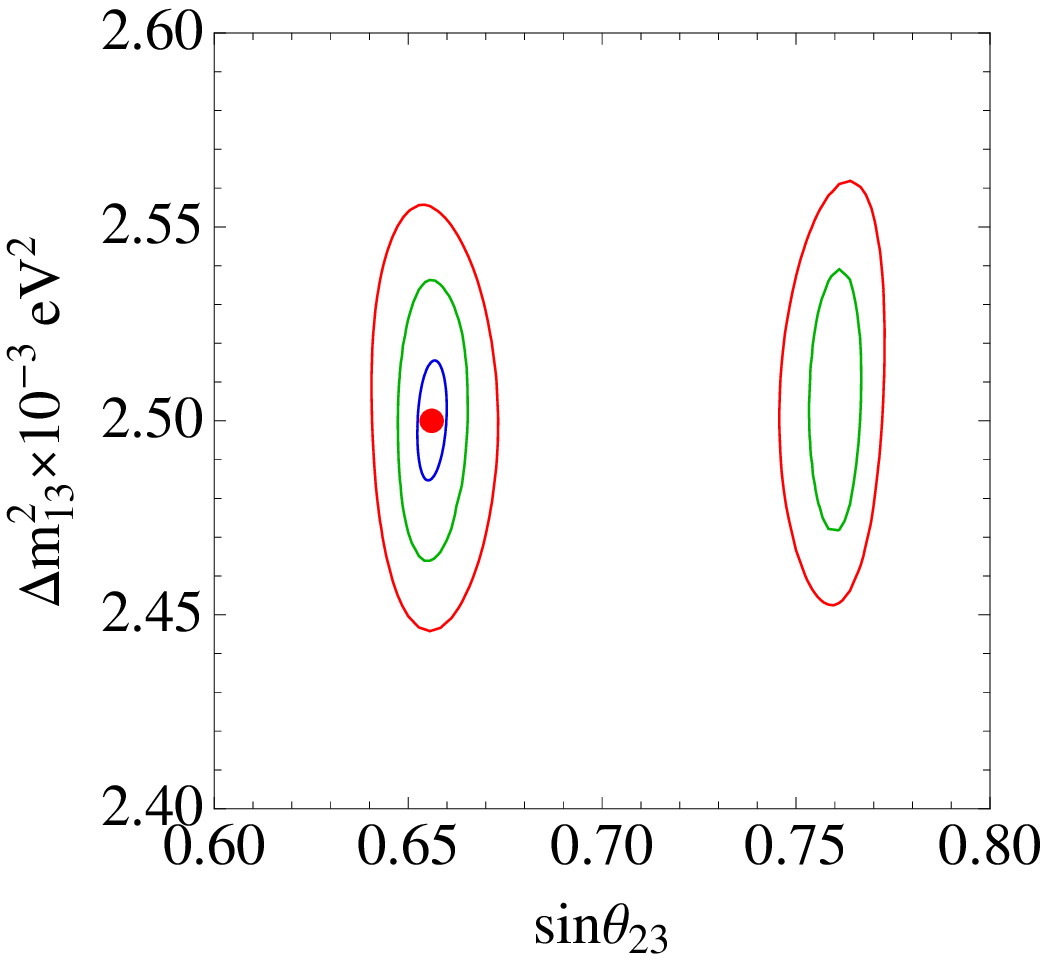}
\caption{Same as Fig.~\ref{fig:contours_sys01} but for $\theta_{13}$ centered around $\sin^22\theta_{13}=0.08$.}
\label{fig:contours_sys08}
\end{figure*}

Figures \ref{fig:contours_sys01} and \ref{fig:contours_sys08} depict the effects of systematic uncertainties, which are often a limiting factor in disappearance experiments. The blue contours assume no systematic error, green correspond to 5\% and red to 10\% systematic errors, for nine energy bins and four angular bins.  In the former two plots we have assumed free $\theta_{13}$ with a gaussian prior centered around the true value $\sin^22\theta_{13}=0.01$ and $\sin^22\theta_{13}=0.08$ respectively, with a 1 sigma uncertainty of 0.02. 

Notice that, while systematic errors from flux/cross-section uncertainties, etc. affect the parameter measurement at some level, they still allow for a fairly precise determination of $\Delta m^2_{31}$ and $\theta_{23}$. As previously discussed, many of the systematics will improve with data, since normalization factors can be included as free parameters in the fit and determined with good precision, especially using the additional energy and angular bins where oscillations are not important. 
Information from near future experiments can also be used to reduce some of the uncertainties.

Good muon track reconstruction would be extremely useful, but relatively precise measurements can be achieved even with reduced angular resolution. The most important factor in the determination of the main oscillation parameters is the energy threshold, as most of the sensitivity comes from neutrino energies between 10 and 40~GeV. With the 10 GeV neutrino energy threshold expected to be reached with the 6 string detector we have considered here and with realistic assumptions about other parameters and uncertainties, the parameter determination will be very good. Two additional strings will be deployed this year in the center of the detector. These will likely improve the quality of the directional reconstruction and the energy threshold and resolution and will thus lead to even better results.

\section{Outlook}
\label{sec:outlook}

The IceCube detector and its Deep Core array provide an excellent 
opportunity for studies of atmospheric neutrinos. Being the largest
existing neutrino detector, it will accumulate a huge number of
atmospheric neutrino events over an enormous energy range, thus
allowing for detailed studies of oscillation physics, Earth density,
atmospheric neutrino fluxes and new physics~\cite{concha0}. In order to extract all this information it is necessary to use energy and angular distribution information, as well as flavor composition, all possible to obtain with the IceCube detector. 

Qualitatively, there are three main energy intervals and three main angular regions which are sensitive to different types of physics.

At very high energies, above $10$~TeV, neutrino interaction
cross-sections become high enough that neutrinos going through the
Earth start being attenuated~\cite{concha1}. 
This effect is sensitive to neutrino interaction cross-sections and to the density profile of the Earth. 

The ``intermediate'' energy region, between $50$~GeV and $1$~TeV can
provide good information about the atmospheric neutrino flux, which
can be used to improve the uncertainties in the simulated atmospheric
neutrino fluxes~\cite{concha2}.

In our paper we concentrated on the ``low'' energy region, below about $40$~GeV, where neutrino oscillation effects can be significant. Although the IceCube detector has higher energy threshold, its low energy extension, the Ice Cube Deep Core array (ICDC), extends the IceCube neutrino detection capabilities in the low energy domain, down to muon or cascade energies as low as $5$~GeV. 

This data can be used for precise measurements of oscillation parameters like $\Delta m^2_{31}$ and $\theta_{23}$ and for resolving some of the neutrino oscillation parameter degeneracies.

A very large effort is directed toward the optimization and construction of a next generation of experiments that will address precision measurements and potential discovery of new phenomena in neutrino oscillations. While they will achieve extremely good sensitivities, these experiments face long construction and data-taking times. In the meantime, ICDC, which is already taking data, will acquire a high statistics atmospheric neutrino sample which can be very useful in extracting information about neutrino oscillation physics. Some of this information could be used for further optimization of the future accelerator experiments. 
Long baseline experiments can achieve very high precision, but have a fixed baseline and a limitted energy coverage. Atmospheric neutrinos in ICDC cover a much more extended range of energies and many baselines and thus provide potentially new, complementary information. Combining data from the two types of experiments would check the consistency of allowed neutrino oscillation parameter space over a large range of energies and propagation distances and can be very important for solving parameter degeneracies. If any new physics is present in the neutrino sector, its effects could be larger at high energies and long distances and would lead to additional effects in the atmospheric neutrino data.

In summary, ICDC offers a unique window toward a better understanding of neutrino properties due to its very high atmospheric neutrino statistics. Careful studies of the expected atmospheric neutrino oscillation signals in ICDC, as the one carried out here, are extremely important, since atmospheric neutrinos constitute an irreducible background to astrophysical neutrino searches and can offer information that is complementary to that obtained from other neutrino experiments.

\section*{Acknowledgments} 

We would like to thank Doug Cowen and Ty DeYoung for useful discussions and suggestions.
This work was supported in part by the NSF grant PHY-0855529. E.~F is supported by the DFG cluster of excellence ``Origin and Structure of the Universeâ" and the European Community under the European Commission Framework Programme 7 Design Study EUROnu, Project Number 212372. O.~M. work is supported by the MICINN (Spain) Ram\'on y Cajal contract, AYA2008-03531 and CSD2007-00060.

\end{document}